\documentclass[aps,prl,reprint,superscriptaddress]{revtex4-2}
\usepackage{amsmath,amsfonts,amssymb,bm,hyperref,graphicx}
\usepackage[notrig]{physics}
\usepackage{CJKutf8}
\usepackage{orcidlink}
\usepackage[final]{changes}
\usepackage{makecell}
\UseRawInputEncoding

\newcommand{\bV}{\bm{V}}

\newcommand{\bT}{\bm{\Theta}}

\begin{document}
\begin{CJK*}{UTF8}{gbsn} % Use default fonts from CJK (see below)

\title{Novel dynamics for an inertial polar tracer in an active bath}
\author{Jing-Bo Zeng (曾敬博)\orcidlink{0009-0003-5926-385X}}
%\email{zengjingbo@stu.pku.edu.cn}
\affiliation{School of Physics, Peking University, Beijing, 100871, China}
\author{Ji-Hui Pei (裴继辉)\orcidlink{0000-0002-3466-4791}}
\email{pjh@stu.pku.edu.cn}
\affiliation{School of Physics, Peking University, Beijing, 100871, China}
\affiliation{Department of Physics and Astronomy, KU Leuven, Leuven, 3000, Belgium}

\begin{abstract}
A polar tracer immersed in an active bath is known to be propelled forward and therefore activated.
Here we report that the induced dynamics of an inertial tracer can be much richer than expected.
We investigate a heavy polar tracer immersed in a bath of independent active Brownian particles.
Using the projection-operator formalism to integrate out the bath, we show that the tracer's reduced dynamics can be \added{precisely }mapped onto a stochastic Lorenz equation. 
According to the attractors in the Lorenz equation, the tracer motion is classified into several different dynamical regimes, 
including active Brownian motion, chiral active Brownian motion, complex chaotic motion, and zigzag active Brownian motion. 
For certain regimes, we derive analytical expressions for the propulsion speed, the velocity covariance, and the effective diffusion coefficient. 
Numerical simulations corroborate these theoretical predictions.
\end{abstract}

\maketitle
\end{CJK*}

\textit{Introduction.}
Active systems consume energy to generate motion. 
From bacteria and microswimmers to artificial robots and flocks of birds, active systems span a wide range of scales. 
Such systems are intrinsically out of equilibrium and exhibit many novel phenomena that are absent in thermal equilibrium. 

Recently, motion of passive objects in active baths has attracted considerable interest, in experiments \cite{Wu2000,Chen2007,Leptos2009,Rafai2010,Kurtuldu2011,Valeriani2011,Mino2011,Mino2013,Kaiser2014,Maggi2014,Patteson2016,Argun2016,Kurihara2017,Maggi2017,Jerez2017,Katuri2021,Paul2022,Boriskovsky2024,Caprini2024,Boudet2025}, theory \cite{Mino2013,Maggi2014,Argun2016,Maggi2017,Katuri2021,Boudet2025,Loi2008,Steffenoni2016,Zakine2017,Burkholder2017,Burkholder2019,Knezevic2020,Kanazawa2020,Solon2022,Granek2022,Feng2023,Ion2023,Sarkar2024,act,Hargus2025,Pei2026}, and simulations \cite{Valeriani2011,Kaiser2014,Paul2022,Caprini2024,Loi2008,Zakine2017,Kanazawa2020,Ion2023,Sarkar2024,Gregoire2001,Underhill2008,Angelani2010,Foffano2012,Mallory2014,Suma2016,Ye2020,Reichert2021,Abbaspour2021,Shea2022,Dor2022,Jayaram2023,Dhar2024,Dolai2024,Singh2024,Shea2024,Kim2024,Sarkar2025}.
% Active systems consume energy to generate motion. 
% From bacteria and microswimmers to artificial robots and flocks of birds, active systems span a wide range of scales. 
% Such systems are intrinsically out of equilibrium and exhibit many novel phenomena that are absent in thermal equilibrium. 
% Recently, the study of active baths has attracted considerable interest \cite{Granek2024,Wu2000,Chen2007,Leptos2009,Rafai2010,Kurtuldu2011,Valeriani2011,Mino2011,Mino2013,Kaiser2014,Maggi2014,Patteson2016,Argun2016,Kurihara2017,Maggi2017, Jerez2017,Katuri2021,Paul2022,Boriskovsky2024,Caprini2024,Gregoire2001,Loi2008,Underhill2008,Angelani2010,Foffano2012,Mallory2014,Suma2016,Knezevic2020,Ye2020,Shea2022,Dor2022,Feng2023,Jayaram2023,Zakine2017,Burkholder2017,Burkholder2019,Kanazawa2020,Reichert2021,Abbaspour2021,Peng2022,Ion2023,Dhar2024,Dolai2024,Zhao2024,Sarkar2024,act,Steffenoni2016,Granek2022,Solon2022,Singh2024,Shea2024,Kim2024,Boudet2025,Sarkar2025,Hargus2025,Pei2026}, aiming to understand how a passive object is influenced by active surroundings. 
Different from the Brownian motion induced by a thermal bath, novel phenomena emerge, 
\replaced{in which}{including the super diffusion for an asymmetric tracer and the negative friction coefficient for a symmetric tracer in 1D. 
In higher dimensions,} the most notable phenomenon is the activation of an anisotropic tracer \cite{Kaiser2014,Angelani2010,Mallory2014,Knezevic2020,Dolai2024,Shea2024}.
Active particles get stuck at confining
walls \cite{Elgeti2013,Schaar2015}, and the local pressure exerted by the active bath depends on the boundary curvature \cite{Takatori2014,Solon2015a}, leading to a nonzero net force acting on an anisotropic tracer. 
Such an effect explains the spontaneous rotation of microscopic gears immersed in active gas and is used to extract work from the active medium \cite{Sokolov2009,Angelani2009,DiLeonardo2010}. 

To fully understand the fluctuation behavior of the passive tracer,
the central theoretical task is to derive the reduced dynamics of the tracer after theoretically integrating out the degrees of freedom of the active bath. 
Let us consider the simplest setup, 
where a 2D polar tracer is immersed in an 
active ideal gas consisting of independent self-propelled particles, 
such as active Brownian particles. 
As one may expect, the reduced dynamics may be characterized by a Brownian motion with an additional net force, associated with the orientation of the tracer. 
Correspondingly, 
the tracer is propelled by the net force and follows dynamics resembling an active Brownian particle (ABP) due to the fluctuation of the orientation. 
Such dynamics has been confirmed for an overdamped tracer both theoretically and numerically \cite{Mallory2014,Knezevic2020,Shea2024}. 
Nevertheless, we would like to show that this picture may completely break down for a tracer with inertia. 

Many previous studies have focused on overdamped tracer because in many situations,
in addition to an active bath, the tracer also experiences a strong friction from the thermal environment.
By contrast, when the tracer is immersed in a purely active environment, such as 
for active baths made of artificial robots at millimeter scale, the inertia of the tracer plays an important role.
%\deleted{Recent study has shown that for an isotropic tracer with soft interaction, the friction may depend on the velocity in a highly nonlinear way \cite{Kim2024,Pei2026}, causing acceleration.}

In this Letter, 
we investigate an underdamped 2D polar tracer with a hardcore interaction.
We report that 
due to the coupling between rotational motion and translational motion, the reduced dynamics of the tracer exhibits rich behaviors beyond the simple directed motion. 
In particular, 
the reduced dynamics is precisely mapped onto a stochastic Lorenz equation. 
Depending on the position of the center of mass and
the ratio between translational and rotational inertia, 
the dynamics is classified into different regimes. 
In addition to the \replaced{ABP regime}{active Brownian motion}, 
\replaced{chiral active Brownian particle (CABP) regime}{chiral active Brownian motion} may emerge from the breaking of symmetry. 
Moreover, the tracer exhibits chaotic motion or enters a zigzag ABP regime in certain cases.
Numerical simulations have verified the above theoretical predictions.

\begin{figure}[t]
\centering
\includegraphics[width=0.9\columnwidth]{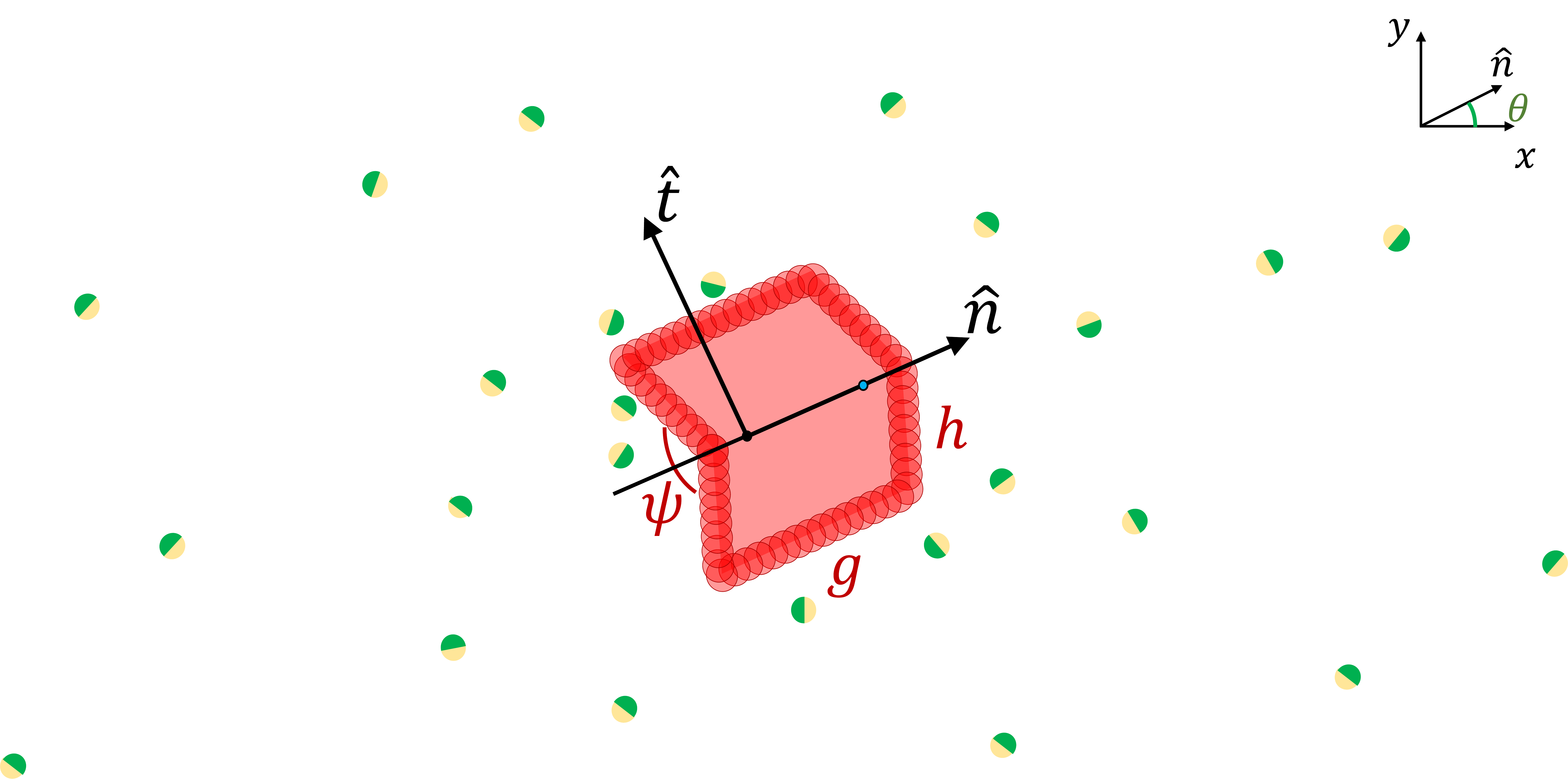}
\caption{
A chevron-shaped rigid tracer (shown in red), composed of a collection of beads and immersed in an active bath, is characterized by its opening angle $\phi$, length $a$, and width $b$.
Each bead is coupled to ABPs via the WCA force.}
\label{fig:schematic}
\end{figure}

\textit{Setup.}
We study a heavy polar tracer in a dilute active bath of $N$ independent overdamped ABPs.
The system is considered in a square domain of side length $L$ with periodic boundary conditions. 
We investigate a chevron-shaped rigid tracer, modeled as a set of beads with fixed relative position, as shown in Fig.~\ref{fig:schematic}. The tracer is described by its center-of-mass position $\bm{r}$ and orientation $\theta$, together with its translational velocity $\bm{v}$ and angular velocity $\omega$. 
Meanwhile, we decompose $\bm{v}$ into longitudinal ($v_n$) and transverse ($v_t$) components, which are respectively parallel and perpendicular to the tracer's polar axis.
The interaction between the tracer and the bath particles is purely collisional; each bead of the tracer interacts with the ABPs via the Weeks-Chandler-Andersen (WCA) force \cite{Weeks1971WCA}.
% \begin{equation}\label{wca}    
% \bm F_c(\bm\eta)=\begin{cases} 0& |\bm\eta|\geq 2^{\frac{1}{6}}a \\ \frac{\epsilon}{a} (\frac{12 a^{13}}{\eta^{13}}-\frac{6a^7}{\eta^7})\hat{\bm\eta} & 0\leq|\bm\eta|\le 2^{\frac{1}{6}}a ,\end{cases}
% \end{equation}
% where $\hat{\bm \eta}$ denotes the unit vector along $\bm \eta$.
The full dynamics of the composite system is given in the End Matter.
We aim at an effective reduced description for the tracer.

\textit{Reduced dynamics.}
Since the tracer is heavy, we can imagine that the motion of the tracer is much slower than the bath particles and there is a time-scale separation.
We mention that a recent study \cite{Kim2024,Pei2026} on spherical tracer reveals that a non-collisional interaction may cause a nonlinear negative friction, 
threatening to the time-scale separation assumption.
Nevertheless, the interaction in the current setup is collisional, so the time-scale separation is ensured as along as the tracer's mass $M$ and moment of inertial $I$ are large.
We apply the projection-operator method to perform the quasi-static expansion and integrate out bath particles. 
The derivation of the reduced dynamics is presented in the Supplemental Material~\cite{[{See }] [{ for detailed theoretical derivations and concrete expressions, which also includes Ref.\cite{Pei2025,Huang2022}}] SM,*null}, in 
which the only approximation is based on the time-scale separation.
Up to the second order in the quasi-static expansion, we obtain the following reduced dynamics for the tracer:
% \deleted{Building on a time-scale separation assumption, previous work \cite{Pei2026} derived the reduced dynamics of a symmetric tracer.} 
% For asymmetric bodies, the coupling between translation and rotation \deleted{due to shape asymmetry }gives rise to new phenomena. Decomposing the tracer's velocity in the \replaced{fixed-body}{body-fixed} frame into longitudinal ($v_n$) and transverse ($v_t$) components \added{(parallel and perpendicular to the tracer’s polar axis)} and applying the projection-operator method, we obtain the corresponding reduced dynamics
\begin{equation}    \label{reduced dynamics}
    \begin{aligned}
    \dot{v}_n&=\omega v_t-\frac{N}{M}(\gamma_{nn}v_n-f_n)+\frac{\sqrt{N}}M\pi_{nn}\xi_n\\
    \dot{v}_t&=-\omega v_n-\frac{N}{M}(\gamma_{tt}v_t+\gamma_{t\theta}\omega)+\frac{\sqrt{N}}{M}(\pi_{tt}\xi_t+\pi_{t\theta}\xi_\theta )\\
    \dot{\omega}&=-\frac{N}{I}(\gamma_{\theta t}v_t+\gamma_{\theta\theta}\omega)+\frac{\sqrt{N}}{I}(\pi_{\theta t}\xi_t+\pi_{\theta\theta}\xi_\theta).
\end{aligned}
\end{equation}
Here $f_n$ is the net force exerted by the active bath; $\gamma$ is the friction matrix while $\gamma_{\theta t},\gamma_{t\theta}$ characterize the coupling between $v_t$ and $\omega$; $\pi$ represents the correlation between stochastic forces; $\xi_n,\xi_t,\xi_\theta$ denote independent Gaussian white noises with zero mean and unit variance. For large $M$ and $I$, the noise can be regarded as weak. The $f_n,\gamma, \pi$ are expressed as expectation values in the bath dynamics with fixed tracer; see Supplemental Material \cite{SM}.
In the following, we use a generalized velocity $\bm u =(v_n,v_t,\omega)$ to denote all three variables. 
We mention that the above dynamics is generally valid for any polar tracer with axial symmetry, not restricted to chevron shape.

The above dynamics defines a nonlinear stochastic differential equation for $\bm u$. It can be precisely mapped onto a stochastic Lorenz equation \cite{Lorenz1963},
\begin{equation}\label{stochastic_Lorenz}
    \begin{aligned}
    \dv{x}{\tau}&=\sigma(y-x)+\xi_1(\tau)\\
    \dv{y}{\tau}&=r x-y-xz+\xi_2(\tau)\\
    \dv{z}{\tau}&=xy-b z+\xi_3(\tau),
    \end{aligned}
\end{equation}
through the following linear change of variables, 
\begin{equation}
    \begin{aligned}
    &\omega = \frac{N\gamma_{tt}}{M}x, \quad 
    v_t = -\frac{N}{M}\frac{\gamma_{\theta\theta}\gamma_{tt}}{\gamma_{\theta t}}y, \\
    &v_n = \frac{f_n}{\gamma_{nn}}-\frac{N}{M}\frac{\gamma_{\theta\theta}\gamma_{tt}}{\gamma_{\theta t}}z, \quad 
    \tau = \frac{N\gamma_{tt}}{M}t.
    \end{aligned}
\end{equation}
% $c_x$, $c_y$, $c_z$, $c_\tau$, $z_0$ in the mapping are constant. 
% Three dimensionless parameters $r$, $\sigma$, and $b$. 
% correlated gaussian noise $\xi_1,\xi_2,\xi_3$ were determinated by the reduced dynamics.
The three dimensionless parameters in the Lorenz equation are given in terms of the parameters in Eq.~\eqref{reduced dynamics},
\begin{equation}
    \sigma = \frac{M\gamma_{\theta\theta}}{I\gamma_{tt}}, \quad 
    b = \frac{\gamma_{nn}}{\gamma_{tt}}, \quad 
    r = \frac{\gamma_{t\theta}\gamma_{\theta t}}{\gamma_{tt}\gamma_{\theta\theta}} + \frac{M f_n \gamma_{\theta t}}{N \gamma_{nn}\gamma_{tt}\gamma_{\theta\theta}}.
\end{equation}
% {The expressions of mapping coefficients ($c_x$, $c_y$, $c_z$, $c_\tau$, $z_0$), 
% three dimensionless parameters in the Lorenz equation ($r$, $\sigma$, $b$), }
See Supplemental Material \cite{SM} for the correlation function of Gaussian noises ($\xi_1,\xi_2,\xi_3$). 

As long as the noise is not too strong (which is fulfilled in the current setup), 
the qualitative features are determined by the corresponding deterministic Lorenz equation. 
The latter exhibits distinct behaviors as the three dimensionless parameters ($r$, $\sigma$, $b$) vary. 
Specifically, depending on the relative magnitude between $r$ 
and $1$, $r_H$, $r_p$ ($r_H$ and $r_p$ are functions of $\sigma,b$),
the dynamical system is controlled by different attractors \cite{sparrowLorenzEquationsBifurcations1982}.
Accordingly, we classify the dynamical regimes of the tracer motion: 
(1) ABP regime: for $r<1$, the system has a unique globally stable fixed point.
(2) CABP regime: for $r>1$ and $r_H<0$ or $1<r<r_H$, the system has a pair of symmetry-related stable fixed points.
(3) Chaotic regime: for $0<r_H<r<r_p$, the system has a strange attractor or periodic orbit in some parameter window.
(4) Zigzag ABP regime: for $r_H>0$ and $r>r_p$, the system has globally stable limit orbit, zigzag ABP motion.

% {
% (1) $r<1$: Unique globally stable fixed point, ABP motion.
% (2) $r>1$ and $r_H<0$ or $1<r<r_H$\deleted{ for $r_H>0$}: A pair of symmetry-related stable fixed points, chiral ABP dynamics. 
% (3) $0<r_H<r<r_p$: Strange attractor or periodic window, chaotic motion.
% (4) $r_H>0$ and $r>r_p$: A global limit orbit, zigzag ABP motion.
% }
% {
% Accordingly, we classify the tracer motion into four dynamical regimes. For $r<1$, the system has a unique global fixed point, yielding ABP motion. For $r>1$ with $r_H<0$ or for $1<r<r_H$ when $r_H>0$, there is a pair of symmetry-related stable fixed points yielding CABP motion. For $r_H>0$ and $r_H<r<r_p$, the system has a strange attractor, or in some narrow period window, stable period orbits giving rise to chaotic motion. For $r_H>0$ and $r>r_p$ system approaches to a global limit cycle, corresponding to zigzag ABP motion.
% }

\begin{table*}[t]
\centering
\begin{tabular}{|c|l|l|l|l|}
\hline
\parbox[c][0.5cm]{4.5cm}{}& \multicolumn{4}{c|}{Mass increasing} \\ 
& \multicolumn{4}{c|}{$\xrightarrow{\hspace{3.5cm}}$} \\
\hline
\parbox[c][0.9cm][c]{4.5cm}{CM located in the front}  & \multicolumn{4}{c|}{\makecell[c]{ABP\\$(r<1)$}} \\ 
\hline
\parbox[c][0.9cm][c]{4.5cm}{CM located at the back; $I/M>\beta_H$} & \makecell[c]{ABP\\$(r<1)$} & \multicolumn{3}{c|}{\makecell{CABP\\$(r>1,\, r_H<0)$}} \\ 
%& $r<1$ & \multicolumn{3}{c|}{$r>1$}\\
\hline
\parbox[c][0.9cm][c]{4.5cm}{CM located at the back; $I/M<\beta_H$} & \makecell[c]{ABP\\$(r<1)$} & \makecell[c]{CABP\\$(1<r<r_H)$} & \makecell[c]{Chaotic\\$(r_H<r<r_p)$} &\parbox{2.8cm}{\makecell[c]{Zigzag ABP\\$(r>r_p)$}} \\
%&$r<1$&$1<r<r_H$&$r_H<r<r_p$&$r_p<r$\\
\hline
\end{tabular}
\caption{Summary of dynamical regimes of the tracer, classified by the CM location and the ratio $I/M$, with the horizontal axis representing increasing mass (while $I/M$ and $N$ fixed).}
\label{tab:your_label}
\end{table*}

% \deleted{The $r$, $\sigma$, $b$ are determined by the parameters of the reduced dynamics.}
We explain below that the tracer's dynamics is mainly determined by the position of the center of mass (CM) as well as the ratios among $M,I,N$.
When the tracer moves in the transverse direction ($v_t$ in Fig.~\ref{fig:schematic}), 
the transverse friction force $f_v$ generally does not act on the CM, 
causing the particles to have a tendency to rotate,
which is characterized by the nonzero $\gamma_{\theta t}$. 
Taking the direction of the propulsion force as a reference, if the CM is located in front of the $f_v$, then $\gamma_{\theta t}<0$; otherwise, $\gamma_{\theta t}>0$.
For $\gamma_{\theta t}<0$, the dynamics is always in the ABP regime. 
For $\gamma_{\theta t}>0$ and small mass, the dynamics is also in the ABP regime.
Keeping $I/M$ and $N$ fixed, as the mass increases, it transitions to chiral ABP behavior. 
For small $I/M$ (less than $\beta_H=\gamma_{\theta\theta}/(\gamma_{nn}+\gamma_{tt})$), a further increase in mass leads to the chaotic regime and eventually to the zigzag ABP regime.
See Tab.~\ref{tab:your_label} for a summary. 
A detailed analysis can be found in Supplemental Material \cite{SM}. 
\begin{figure*}[!ht]
    \centering
    \includegraphics[width=2\columnwidth]{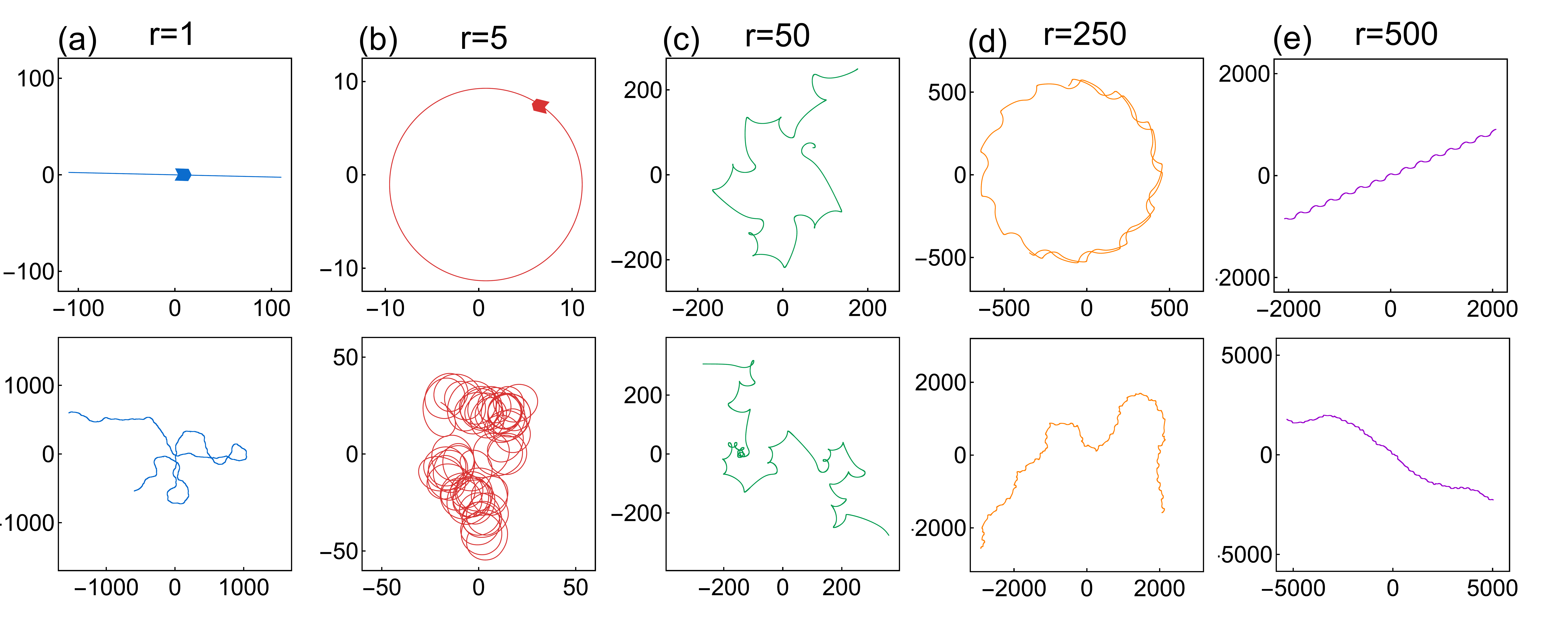}
    \caption{Typical trajectories of CM from the reduced dynamics in different dynamical regimes: (a) ABP regime; (b) CABP regime; (c)(d) chaotic regime; (e) zigzag ABP. 
    In each column, the figure in the bottom displays trajectories in the presence of noise, while the figure in the top depicts trajectories in the zero-noise limit.
    As the column changes, $r$ varies, while $\sigma=10.6$ and $b=2.95$ are fixed.}
    % \caption{\replaced{Figures in }{In Figs. (a-e), }the top row display\deleted{s} representative tracer trajectories for varying $r$ at fixed $\sigma$ and $b$. \replaced{Figures in t}{T}he bottom row depict\deleted{s} the corresponding deterministic trajectories (where noise amplitude is zero). As $r$ varies, the tracer exhibits distinct regimes of motion: (a) ABP, (b) CABP, (c,d) chaotic, and (e) zigzag ABP.}
    \label{fig2}
\end{figure*}

\textit{Different dynamical regimes.}
In the following, 
we analyze the behaviors in different regimes in detail. 

(1) ABP dynamics.
For $r<1$, the Lorenz dynamics has a globally stable fixed point $C_0$ at $(v_n,v_t,\omega)=(f_n/\gamma_{nn},0,0)$. In the absence of noise, the tracer moves at constant velocity along the direction of propulsion. With weak noise, the heading is perturbed, so the tracer behaves like an ABP.
For both zero noise and finite noise cases, we show the typical trajectory of the tracer in Fig.~\ref{fig2}(a).

(2) CABP dynamics.
For $r>1$ with $r_H<0$ or $1<r<r_H$, 
the Lorenz equation has a pair of symmetry-related fixed points $C_\pm$ at $\bm u=\pm(v_{n0},v_{t0},-\omega_{0})$, with $v_{n0}=N(\gamma_{\theta\theta}\gamma_{tt}-\gamma_{\theta t}\gamma_{t\theta})/(M{\gamma_{\theta t}})$, $v_{t0}=\sqrt{N\gamma_{\theta\theta}(f_n-\gamma_{nn}v_{n0})/(M{\gamma_{\theta t}})}$ and $\omega_0=\sqrt{N\gamma_{\theta t}(f_n-\gamma_{nn}v_{n0})/(M{\gamma_{\theta\theta}})}$.
In the absence of noise, the tracer follows uniform circular motion with speed $v_p=\sqrt{v_{n0}^2+v_{t0}^2}$ and radius $R=v_p/\omega_0$. 
The spin and the orbital motion are locked, possessing the same period $2\pi/\omega_0$.
Two different fixed points respectively correspond to the clockwise and counter-clockwise motions.
The tracer velocity and the propulsion force are not aligned, with angle difference $\arctan(v_{t0}/v_{n0})$.
For weak noise, the direction of the propulsion force is perturbed so that the tracer no longer moves along a closed circle but behaves as a CABP.
Typical trajectories are shown in Fig.~\ref{fig2}(b).

In the presence of weak noise, there are transitions between the clockwise motion and the counter-clockwise motion, but the transitions are rare. 
The chirality of the motion is maintained in a fairly long time. 
Recall that our original composite system of tracer plus bath respects the full chiral symmetry, indicating that this induced active motion arises from a spontaneous breaking of the chiral symmetry \cite{li_spontaneous_2017}.
It reveals the possibility of emergent chiral active motion from non-chiral ones.

(3) Chaotic motion. When $1<r_H<r<r_p$, in most of the parameter regime, the Lorenz equation has a strange attractor which causes a chaotic motion even if there is no noise, as shown in Figs.~\ref{fig2}(c) and the supplemental videos. 
In this case, the trajectory of $\bm u$ in the Lorenz equation exhibits a famous butterfly pattern \cite{Strogatz2018}. 
Typically, when $\bm u$ transits between two wings of the butterfly attractor, the tracer exhibits a long displacement in the position space, 
and when $(v_n,v_t,\omega)$ hovers on one side of the wing, 
the tracer moves slowly but spins more. 
The motion of the dynamics can be roughly summarized as chaotic combinations of the above two behaviors.

In some narrow parameter windows, the Lorenz equation may have doubling stable periodic orbits, but these orbits are fragile under noise.
The system is easily driven into transient chaos by noise. 
Therefore, in the absence of noise the tracer exhibits a quasi-periodic circular motion, as shown in Fig.~\ref{fig2}(d). 
In the presence of noise, the trajectory of the tracer becomes similar to the previous chaotic case due to the noise induced chaos.
%Let us regard the noiseless trajectory as a tightly wound rope, and as noise grows it gradually unravels and gets jumbled. 
%\added{This is the so-called noise-induced chaos which may also occur in the CABP regime when the noise is sufficiently large \cite{Huang2022}.}

(4) Zigzag ABP. When $r_H>0$ and $r>r_p$, the Lorenz equation possesses a globally stable periodic orbit, which is symmetric with respect to the $v_n$ axis. 
In the case of zero noise, as $\bm u$ evolves along this periodic orbit for one cycle, the tracer undergoes a net forward displacement in position space, while its orientation angle returns to its original value after one oscillation period.
%{correspondingly in the position space, the tracer undergoes a net forward displacement while the angular motion recovers the original value after a period of oscillation.}
Then, in subsequent cycles, the tracer moves forward in the same direction.
Therefore, the tracer exhibits an overall zigzag motion like a swing car. 
In the presence of noise, despite the short-time zigzag motion, the long-time behavior resembles an ABP. 
Please see trajectories in Fig.~\ref{fig2}(e), as well as supplemental videos. 

\textit{Diffusion coefficient.} 
% In the long-time limit, the tracer motion is diffusive, 
% with mean-square displacement $\ev{ r^2(t)}=4Dt$, 
% where $D$ is the diffusion coefficient. 
Active motions may enhance diffusion \cite{Wu2000, Leptos2009}. 
Here we find that the diffusion coefficient $D$ exhibits distinct behaviors across different dynamical regimes.

% For the traditional Brownian motion, the tracer's diffusion coefficient 与质量无关, however for tracer in active bath the tracer undergoes active motion, its diffusion is enhanced by active motion\cite{Wu2000,Leptos2009}, the diffusion coefficient is influenced by mass and we will further show that the dependence will behave different for different regime
% Since the tracer undergoes active motion, we expect its diffusion to be enhanced \cite{Wu2000,Leptos2009}. In fact, we further observe that diffusion behaviors are distinct across different dynamical regimes. 
\begin{figure}[!ht]
    \centering
    \includegraphics[width=1\linewidth]{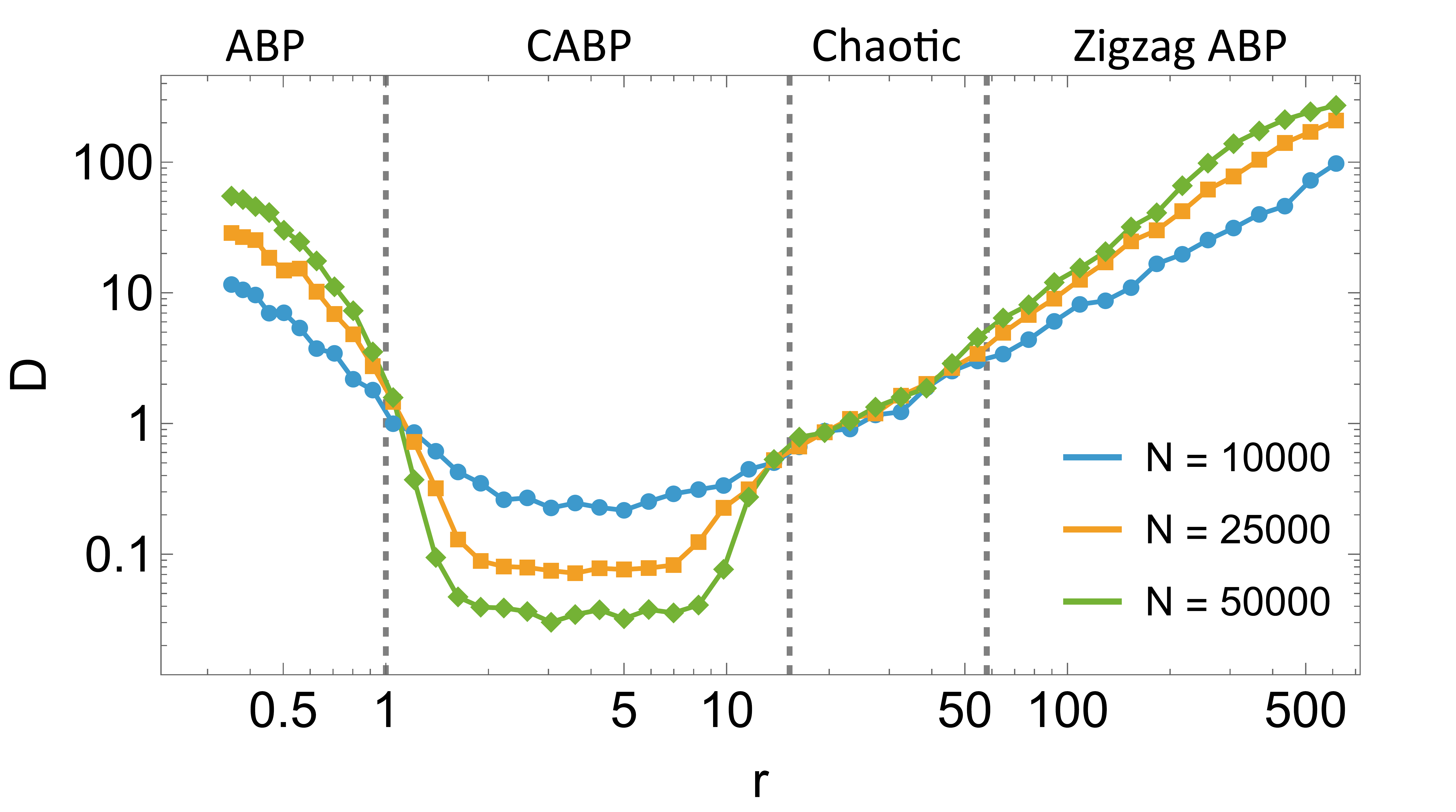}
    \caption{The dependence of the diffusion coefficient on $r$, with fixed $\sigma=4.5$ and $b=1$.
    %\added{As $r$ increases, the dynamical regimes change }
    For a given $r$, a larger number of bath particles $N$ corresponds to a lower noise strength, depicted in different colors.}
    \label{fig:diffisionCoefficient}
\end{figure}

We plot the diffusion coefficient obtained from simulations of the reduced dynamics in Fig.~\ref{fig:diffisionCoefficient}. 
Different dynamical regimes are realized by varying \(r\) (horizontal axis) while keeping \(\sigma\) and \(b\) fixed, which corresponds to varying the mass \(M\) while keeping \(I/M\) fixed at a small value. For each fixed \(r\), lower noise amplitudes are achieved by increasing the number of bath particles, as indicated by the different colors.

% We plot the diffusion coefficient obtained from the simulation of reduced dynamics, for different dynamical regimes in Fig.~\ref{fig:diffisionCoefficient}.
% We realize different dynamical regimes by varying $r$ (horizontal axis) with fixed $\sigma$ and $b$, corresponding to vary mass $M$ while $I/M$ fixed to be small. For fixed $r$, we realize lower noise amplitudes by increasing the number of the bath particles, shown in different colors. 
% In contrast to conventional Brownian motion where the diffusion coefficient is independent of the mass, 
% here tracer's mass is directly related to different dynamical regimes and therefore influences the diffusion behavior significantly. 

In contrast to conventional Brownian motion, where the diffusion coefficient is independent of the particle mass, here the tracer mass is directly tied to the dynamical regime and therefore has a significant influence on the diffusion behavior.
In the (zigzag) ABP regime, active motion enhances diffusion, and a lower noise amplitude increases both the persistence length and diffusion coefficient.
In the CABP regime, active motion fails to enhance diffusion, and the diffusion is purely induced by noise. 
Therefore, a lower noise strength decreases the diffusion, and the diffusion coefficient exhibits a low plateau while $r$.
In the chaotic regime, active motion itself becomes diffusive in nature, leading to nearly identical diffusion coefficients across different noise strengths.

\textit{Linear approximation.}
In the ABP and CABP regimes, the Lorenz equation is attracted by a stable fixed point. 
Thus, we may employ a linear approximation for $\bm u=(v_n,v_t,\omega)$ near the fixed point (precisely speaking, near the expectation value; see below). 
%Precisely speaking, the expansion is around the expectation value of $\bm u$, which deviates from the fixed point due to the nonlinearity of Lorenz equation. 
The dynamics is approximated by an Ornstein-Uhlenbeck (OU) process,
\begin{equation}\label{OU}
    \dot{\bm u}=-\bm{\Theta}\cdot(\bm{u}-\bm\mu)+\delta\bm{F},
\end{equation}
where $\bT$ is the drift matrix, and  $\delta\bm{F}$ is Gaussian noise, with correlation matrix denoted by $\ev{\delta\bm{F}(t)\delta \bm{F}^T(0)}=2\bm{B}\delta(t)$.
%We denote the noise correlation as $\bm{B}=\langle\delta\bm{F}\delta \bm{F}^T\rangle$ and the velocity fluctuation as $\delta\bm u=\bm u-\bm\mu$. The velocity $\bv$ has a mean value of $\bmu$ and its covariance matrix $\bm{V}=\langle\delta\bv\delta\bm u^T\rangle$. 

Since the drift in Eq.~\eqref{reduced dynamics} is nonlinear, the mean value of $\bm u$ generally does not coincide with the deterministic fixed point.
A direct Taylor expansion around the deterministic fixed point may be inaccurate. 
To construct a more accurate equation, coefficients ($\bm\mu$, $\bm{\Theta}$ and $\bm{B}$) are determined such that the mean value and the covariance matrix in steady state coincide with those from the reduced dynamics (precisely speaking, equivalent linearization method \cite{sochaEquivalentLinearizationStochastic2008}).
To calculate the expectation values and the covariance matrix for the reduced dynamics,
we adopt the moment-closure method \cite{kuehnMomentClosureBrief2016}, see Supplemental Material \cite{SM}
for details. 
%\added{Up to the third order of the moment closure}
% We utilize the expectation values to linearize the deterministic part of the reduced dynamics, thereby determining $\bm{\Theta}$ and $\bm{\mu}$.
% Finally, we determinate the noise matrix $\bm{B}$ through Lyapunov equation $\bm{B=1/2(\bm{\Theta}\bm{V}+\bm{V}\bm{\Theta}^T)}$. 
By using the OU dynamics \eqref{OU}, we can deduce the time-dependent velocity correlation matrix $\bV(t)=\langle\delta\bm u(t)\delta\bm u(0)\rangle$.
From this, we may further obtain the analytical expression of the mean-squared angular displacement $\langle \delta\theta(t)^2\rangle$, as well as the diffusion coefficient through $D=1/2 \int_0^\infty\dd t \langle v_x(0)v_x(t)+v_y(0)v_y(t)\rangle$\added{, where $v_x$ and $v_y$ are the tracer velocity components in the laboratory frame }\cite{Teeffelen2008,scholz_inertial_2018}.
Explicit expressions are provided in the Supplemental Material \cite{SM}.
%Do the calculate for expectation we get
% \begin{align}
%     D&=\int_0^\infty\dd{s}\frac12e^{-\frac 12\langle\delta\theta^2(s)\rangle}[\cos(\mu_\theta s)(\mu_0^2+V_{nn}(s)+V_{tt}(s)-\notag\\&V_{n\theta}^2(s)-V_{t\theta}^2(s))-\sin(\mu_\theta s)(\mu_nV_{n\theta}(s)+\mu_tV_{t\theta}(s))].
% \end{align}
%and correlation between angle and velocity $V_{n\theta}(t)=\langle\delta v_n(0)\delta\theta(t)\rangle,V_{t\theta}(t)=\langle\delta v_t(0)\delta\theta(t)\rangle$.
% \begin{align}
%     \langle\delta\theta(t)\delta\theta(t)\rangle=2\int_0^t\dd t'\int_0^{t'}\dd sV_{\omega\omega}(s)\notag\\
%     \bm{V}_{n\theta}(t)=\langle \delta v_n(0)\delta \int_0^t\dd s\omega(s)\rangle=\int_0^t\dd sV_{n\omega}(s)\notag\\
%     \bm{V}_{t\theta}(t)=\langle \delta v_t(0)\delta \int_0^t\dd s\omega(s)\rangle=\int_0^t\dd sV_{t\omega}(s),
% \end{align}
% here $V_{n\omega},V_{t\omega},V_{\omega\omega}$ are the components of correlation function. We then calculate mean-squared displacement (MSD) via
% \begin{align}
%     \langle r^2(t)\rangle=2\int_0^t\dd t'\int_0^{t'}\dd s \langle v_x(0)v_x(s)+v_y(0)v_y(s)\rangle.
% \end{align}
%Here the $v_x$ equals to $v_n\cos\theta-v_t\sin\theta$ and $v_y$ equals to $v_n\sin\theta+v_t\cos\theta$. At short times the motion is ballistic $\langle r^2(t)\rangle\propto t^2$ with self-propulsion velocity $\mu_0=\sqrt{\mu_n^2+\mu_t^2}$, while at long times it crosses to diffusive behavior, 

\textit{Numerical confirmation.}
To validate our reduced dynamics, we perform numerical simulations of the original composite system of tracer plus bath. Here we focus on the comparison between the composite simulation and the reduced dynamics in the ABP and CABP regimes. In these two regimes, the linear approximation provides analytical results for the reduced dynamics.

\begin{figure}[!ht]
    \centering
    \includegraphics[width=\linewidth]{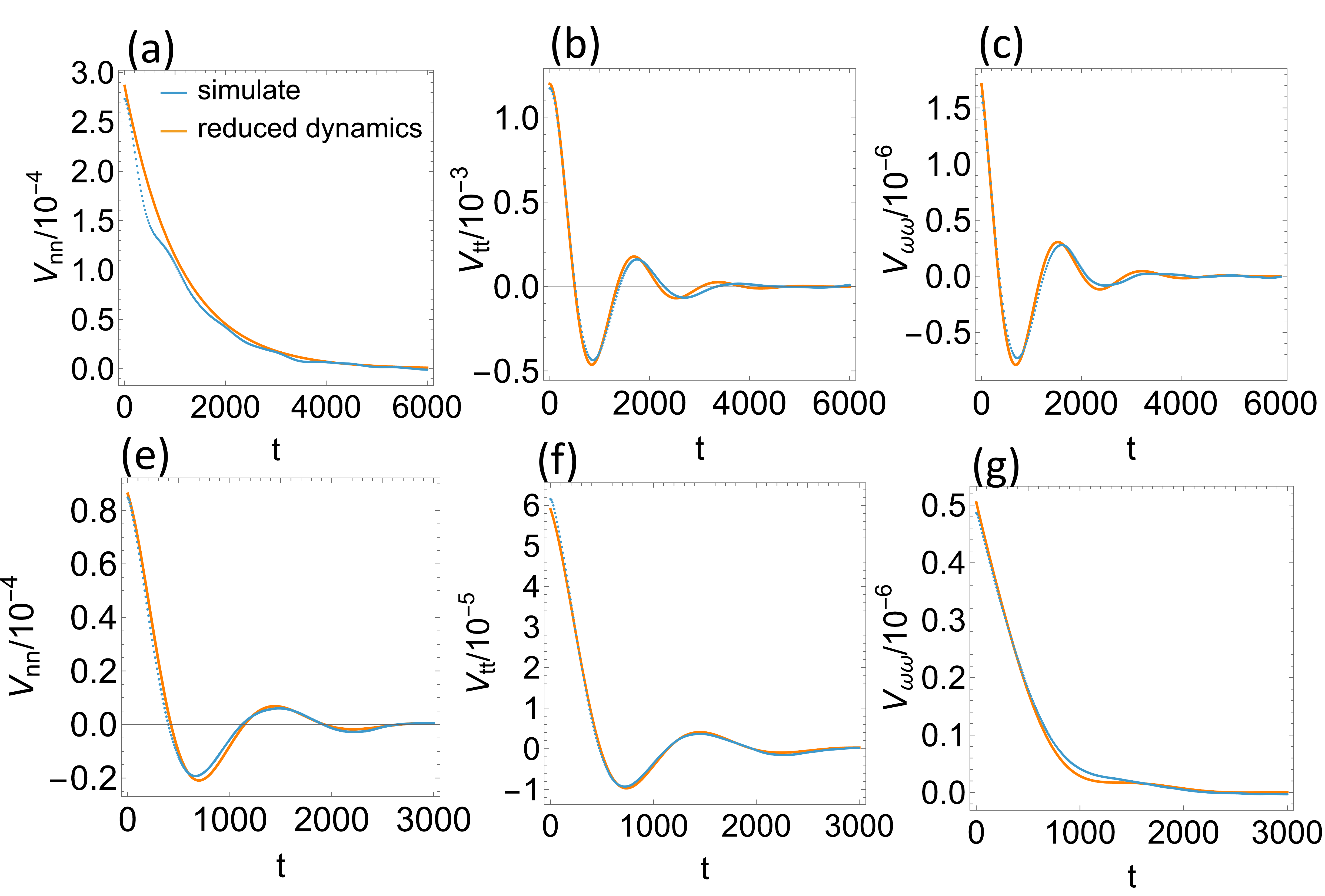}
    \caption{Comparison of the velocity autocorrelation functions between linear approximation of the reduced dynamics (shown in orange) and simulations of composite system (shown in blue). (a)-(c) shows for the ABP regime, (d)-(f) are for the CABP-regime.}
    \label{fig:correlation}
\end{figure}

We calculate the time-dependent autocorrelation function of velocity $V_{ii}(t)$ ($i=n,t,\omega$) from the simulation of the composite system and compare it with the analytical prediction from the linear approximation. The results are presented in Fig.~\ref{fig:correlation}.
The excellent agreement between the analytical curves (orange solid lines) and the simulation data (blue dots) confirms that the reduced dynamics provides an effective description of the tracer motion.

\textit{Conclusion.}
We derive the reduced dynamics for an inertial polar tracer in an active bath via the projection-operator formalism 
and demonstrate that it can be mapped onto a stochastic Lorenz equation. This central finding reveals a remarkably rich dynamical landscape controlled by the tracer's mass and shape. We classify the tracer motion into distinct regimes: ABP motion, CABP motion, chaotic motion, and zigzag ABP motion. 
In particular, the existence of CABP regime indicates that chiral active motion can be induced from non-chiral active environment from spontaneous symmetry breaking.

The diffusion coefficient differs substantially across the different regimes.
For the ABP and CABP regimes, we develop an effective linear approximation to capture the average and velocity correlation functions.
These theoretical predictions were validated by numerical simulations of composite system.

Our findings establish a direct bridge between the dynamics of active matter and the paradigms of chaos theory.
This work demonstrates that a tracer's transport mode can be precisely controlled by tailoring its physical properties, opening new avenues for designing microswimmers and understanding transport in complex active environments.

\begin{acknowledgments}
    \textit{Acknowledgments. }The authors thank H. T. Quan and C. Maes for helpful comments on the manuscript. This work is supported by the National Natural Science Foundation of China (NSFC) under Grants No. 12375028, and No. 12521004.
\end{acknowledgments}

\bibliography{main_new}

\section*{End Matter}
Here we specify the microscopic dynamics of the full composite system.
The system consists of an inertial chevron-shaped rigid tracer immersed in a dilute bath of N independent ABPs. The tracer is characterized by its center of mass position $\bm r$, orientation $\theta$, translational velocity $\bm v$, and angular velocity $\omega$.
Each bath particle is described by its position $\bm\eta^a$ and propulsion angle $\phi^a$ $(a=1,\dots, N)$. 

%\textcolor{orange}{should clarify the meaning of fixed-body frame position, is it the position relative to the CM?}
The tracer is modeled as a rigid collection of $n_b$ beads. We use $\bm \rho_i$ ($i=1,\dots ,n_b$) to denote the position of bead $i$ relative to the tracer center of mass in the body-fixed frame. Accordingly, considering the orientation of the tracer, the position of bead $i$ in the laboratory frame is 
\begin{align}
    \bm r_i=\bm r+\bm R(\theta)\bm\rho_i,
\end{align}
where $\bm R(\theta)$ is the two-dimensional rotation matrix.

Each bath particle interacts with every tracer bead through a Weeks-Chandler-Andersen (WCA) force,
\begin{equation}
\bm F_c(\bm s)=
\begin{cases}
\displaystyle
\frac{\epsilon_c}{\sigma_c}
\left[
12\left(\frac{\sigma_c}{|\bm s|}\right)^{13}
-6\left(\frac{\sigma_c}{|\bm s|}\right)^7
\right]\hat{\bm s},
& |\bm s|<2^{1/6}\sigma_c, \\[2mm]
0, & |\bm s|\ge 2^{1/6}\sigma_c,
\end{cases}
\end{equation}
where $\hat{\bm s}=\bm s/\abs{\bm s}$ denotes the direction. Parameter $\epsilon_c$ sets the interaction strength, and $\sigma_c$ is the interaction length scale.

The total force each bath particle exerted on the tracer can be written as
\begin{equation}
\bm F(\bm r,\theta;\bm\eta)
=-\sum_{i=1}^{n_b}\bm F_c\!\left(\bm\eta-\bm r-\bm R(\theta)\bm \rho_i\right),
\end{equation}
while the corresponding torque about the tracer center of mass is given by
\begin{equation}
\Gamma(\bm r,\theta;\bm\eta)
=-\sum_{i=1}^{n_b}
\left[\bm R(\theta)\bm \rho_i\right]\times
\bm F_c\!\left(\bm\eta-\bm r-\bm R(\theta)\bm \rho_i\right)\cdot \hat{\bm z}.
\end{equation}
We have set the counter-clockwise rotation as the positive direction for the torque.
 
According to the theory for a 2D rigid body,
the coupled equations of motion are given by
\begin{equation}
    \label{eq:full_dynamics}
    \begin{aligned}
    \dot{\bm{\eta}}^a &= v_0\,\bm{e}(\phi^a)-\chi\,\bm{F}(\bm{r},\theta;\bm{\eta}^a),\qquad \bm{e}(\phi)=(\cos\phi,\sin\phi),\\
    \dot{\phi}^a &= \sqrt{2D_r}\,\xi^a(t),\\
    M\dot{\bm{v}} &= \sum_{a=1}^N \bm{F}(\bm{r},\theta;\bm{\eta}^a),\qquad \dot{\bm{r}}=\bm{v},\\
    I\dot{\omega} &= \sum_{a=1}^N \Gamma(\bm{r},\theta;\bm{\eta}^a),\qquad \dot{\theta}=\omega.
    \end{aligned}
\end{equation}
Here, $v_0$ is the self-propulsion speed of the ABPs, $\chi$ is their mobility, and $D_r$ is the rotational diffusion coefficient. The noises $\xi^a$ are independent Gaussian white noises satisfying
$    \ev{\xi^a(t)\xi^b(t')}=\delta_{ab}\delta(t-t')$.

Equation~\eqref{eq:full_dynamics} defines the original composite dynamics in our study. 
%It is used in our direct numerical simulations. 
Starting from this microscopic description, 
the reduced tracer dynamics in Eq.~\eqref{reduced dynamics} is obtained by assuming a separation of time scales. 

The bath-particle mobility is set to $\chi=0.02$, the propulsion velocity is set to $1$, and the angular diffusion is set to $0.04$.
The tracer is specified by $h=7$, $g=12$, and $\psi=120^\circ$, with \(66\) interaction points uniformly distributed along the boundary with spacing \(0.8\). 
In the WCA interaction, the strength is $\epsilon=1/6$ and the interaction length scale is taken as $\sigma_c=2^{-1/6}$.

% We provide supplemental videos for both the direct simulation of the composite tracer--bath system and the reduced dynamics.
% For the direct composite-system simulations, the videos illustrate the ABP, CABP, and chaotic regimes. 
In the supplemental videos, the center of mass is chosen to be located at the back of the tracer, 
on the symmetry axis at the rear cusp of the double-V shape.

\end{document}